\documentclass[reprint,prl,superscriptaddress]{revtex4-1}
\usepackage{graphicx}
\usepackage{subfigure}
\usepackage{dcolumn}
\usepackage[T1]{fontenc}
\usepackage{mathptmx}
\usepackage{diagbox}
\DeclareGraphicsExtensions{.pdf,.jpg,.png}

\begin{document}

\title{Mitigating Congestion in Complex Transportation Networks via Maximum Entropy}

\author{Yuhang Fan}
\affiliation{State Key Lab. of Industrial Control Technology, Zhejiang University, Hangzhou, 310027, China}
\affiliation{School of Mathematics, Zhejiang University, Hangzhou, 310027, China}
\author{Hanyuan Liu}
\affiliation{College of Information Science and Electronic Engineering, Zhejiang University, Hangzhou, 310027, China}
\author{Shibo He}
\affiliation{State Key Lab. of Industrial Control Technology, Zhejiang University, Hangzhou, 310027, China}

\begin{abstract}
In this paper, we reveal the relationship between entropy rate and the congestion in complex network and solve it analytically for special cases. Finding maximizing entropy rate will lead to an improvement of traffic efficiency, we propose a method to mitigate congestion by allocating limited traffic capacity to the nodes in network rationally. Different from former strategies, our method only requires local and observable information of network, and is low-cost and widely applicable in practice. In the simulation of the phase transition for various network models, our method performs well in mitigating congestion both locally and globally. By comparison, we also uncover the deficiency of former degree-biased approaches. Owing to the rapid development of transportation networks, our method may be helpful for modern society.
\end{abstract}
\maketitle
Entropy, as a key concept, plays an important role in many fields such as information theory, statistical physics and so on \cite{shannon2001mathematical,reichl1980modern}. In the study of complex network, the entropy has been used to characterize the properties of network topology \cite{rosvall2005networks,bianconi2009entropy} and diffusion process on network \cite{gomez2008entropy}. Via application of maximum entropy method, we can optimize the network topology for special function and make prediction of network properties \cite{park2004statistical,burda2009localization}.

Traffic flow, modeling the process of material exchange between cells, packet transfer on the Internet, municipal traffic and so on \cite{arenas2001communication,li2015percolation,sole2016congestion}, is one of the significant research fields in complex networks. While traffic congestion is the main hindrance to transport network's normal and efficient functioning \cite{zhao2005onset}, some models, such as random walk \cite{eisler2005random}, shortest path \cite{goh2001universal} and so on, has been built to analyze this phenomenon and studies has been focused on optimizing network topology and improving the traffic capacity to achieve free-flow for these different models \cite{liu2007method,guimera2002optimal}. Most of previous studies concentrate on optimizing the topology of network and designing better routing strategies \cite{kleinberg2000navigation} to improve the efficiency of transportation. These methods, usually needing high cost to change infrastructure or very concrete and global information of the traffic process, are designed for some special cases and maybe uneconomical and unattainable in practice.

Our study is complementary to the above efforts. Here, based on maximum entropy method, we develop an approach to mitigating congestion by allocating limited capacity to nodes reasonably in network with observable information in traffic process. We reveal that we can improve the whole system's traffic efficiency by just improving a little node capacity without changing network topology. Our method is tested in various models for different networks, performing well in raising traffic efficiency.

Before presenting our approach, we give a brief description of network traffic. At each time step, each node has a probability to generate a packet which has a destination and chooses a path towards its destination in network immediately. And a node $N_i$ can deliver at most $C_i$ packets to the next node of its path at a time step. As the capacity is limited, the packets which arrive to the node earlier have priority to be delivered. A packet is removed from the traffic after reaching its destination.

We suppose that we can get each node's local information including the frequency of different neighbors it delivers packets to and the number of packets it delivers in traffic. The frequency and number, as local quantities decided both by the traffic dynamical process and by the network topology, is typically easy to be observed and counted in reality. Let us analyze the routing of packets which is certain under specific strategy from a stochastic standpoint. Mathematically speaking, we regard the traffic process on a network composed from $n$ nodes as a time invariant \emph{ergodic Markov chain} with a transition matrix $M=(q_{ij})$, where the entry $q_{ij}$ is the average frequency of packets to go from node $N_i$ to $N_j$ at a time step satisfying $\sum\nolimits_{j=1}^nq_{ij}=1\forall i$. We introduce entropy rate to characterize this process. An ergodic Markov chain with transition matrix $M=(q_{ij})$ has a unique stationary distribution $w=(w_1,w_2,\dots,w_n)^T$ satisfying
\begin{equation}\label{eq:station}
  M^Tw=w,
\end{equation}
where $M^T$ is the transposed matrix of $M$. And $w$ can be computed directly with normalization condition $\sum\nolimits_{j=1}^nw_i=1$. The dynamical properties of the network can be evaluated by entropy rate $s$ that is given by
\begin{equation}\label{eq:entropy}
  s=-\sum_{i,j}w_iq_{ij}\ln q_{ij}.
\end{equation}
The entropy rate is a quantity studied in the diffusion process and measure the time density of the average information in a stochastic process. It has been showed that a high entropy rate indicates a large randomness, or easiness of propagating from one node to another \cite{gomez2008entropy,burda2009localization,berger1996maximum}. Therefore it can also be related to an efficient and free traffic over the network. In that paper, our main goal is to mitigating traffic congestion by maximizing the associated entropy rate function.

A very simple method to mitigate congestion is to improve the capacity $C_i$ of each node $N_i$ as high as possible. However, the capacity is usually limited by various conditions. And an fundamental and significant problem is about how to allocate the limited capacity to a given network to make the system efficient. Next we try to solve this problem by maximizing the entropy rate. The node $N_i$ in network corresponding to the row vector $v_i=(q_{i1},q_{i2},\dots,q_{in})$ of network's transition matrix $M=(q_{ij})$, where $q_{ij}$, according to the above generation method of matrix $M$, can represent the possibility of a packet at node $N_i$ going to node $N_j$ at a time step in large scale approximately. When congestion occurs in node $N_i$, we have $q_{ii}>0$, that is, the packets have a possibility to wait in node $N_i$ for being delivered at a latter time step. And the direct consequence of improving the node capacity $C_i$ is the decrease of $q_{ii}$. We denote the $q_{ii}$ as the \emph{detention probability} of node $N_i$ in this paper, which plays an important role in our model. Suppose the change of $q_{ii}$ is $p$ ($0\leq p_i\leq q_{ii}$) after being allocated new capacity (i.e. the probability of a packet stay in the node $N_i$ is $q_{ii}-p_i$). Since the nodes deliver the packets by time-priority which is not associated with the packets' routing, it is natural to get that the $j$th ($j\neq i$) component of vector $v_i$ is $(1+{p_i}/(1-q_{ii}))q_{ij}$. Therefore we can view $v_i$ as a vector function of variables $p_i$. While $M$ is composed from $\{v_i\}_{1\leq i\leq n}$, we can get the entropy rate $s$ as a function of variable $p=(p_1,\dots,p_n)$ from equations \ref{eq:station} and \ref{eq:entropy}.

It is obvious that a higher node capacity $C_i$ will lead to a lower detention probability $q_{ii}$. And we need a quantitative relationship to identify the feasible region of $(p_1,\dots,p_n)$. As mentioned above, we suppose we can get the the average number of packets that the node $N_i$ received at each time step is $Q_i$, which can be counted relatively easily in practice or be computed analytically for some special cases \cite{zhao2005onset}. Suppose the change of $C_i$ is $\mathrm{d}C_i$. Approximately, we have that
\begin{equation}\label{eq:piCi}
  p_i=-\mathrm{d}q_{ii}\approx \frac{\mathrm{d}C_i}{Q_i},
\end{equation}
that is, the decrease of detention probability approximately equal to the ratio of the increment of the number of packets a node can deliver per time step and the the average number of packets that the node $N_i$ received at each time step. Then for a given model and limited node capacity budget $B$. we model our problem as follow: for
\begin{eqnarray}\label{eq:feasible}
  \sum_{i=1}^{n}p_iQ_i&=&\sum_{i=1}^{n}\mathrm{d}C_i\leq B, \\
\nonumber  0\leq &p_i&\leq q_{ii},\quad \forall i,
\end{eqnarray}
maximize the function $h(p_1,\dots,p_n)$
\begin{equation}\label{eq:maxh}
  \max h(p_1,\dots,p_n).
\end{equation}
From equations \ref{eq:feasible}, the feasible region of optimization variable $p$ is a convex set. And the entropy function have some nice properties. Many sophisticated methods, such as Monte Carlo algorithms \cite{bohn2007structure}, have been developed from such problem before \cite{boyd2004convex}, and can be applied directly. Suppose that $h$ get its maximum at $p^*=(p_1^*,\dots,p_n^*)$ on the feasible region. We allocate $\mathrm{d}C_i=p_i^*Q_i$ capacity to node $N_i$. Generally, $p_i^*Q_i$ is not integer and we need to adjust our distribution as close as possible to it. For the case the function $h$ is analytic at $0$,
\begin{equation}\label{hpartial}
  h(p)=h(0)+\sum_{i=1}^{n}\frac{\partial h}{\partial p_i}\vert_{p=0}p_i+o(\|p\|).
\end{equation}
By the method of Lagrange multipliers, we can range nodes in network by the priority $h_{p_i}(0)/Q_i$ for the case of limited traffic capacity budget approximately, and allocate the budget to the nodes with higher priority preferentially within the feasible region. Incidently, with less information, since
\begin{equation}\label{limw}
  \lim_{t\to\infty}M^{Tt}x=w
\end{equation}
for any initial distribution $x$, $Q_i$ can be approximated by $Qw_i$, where $Q$ is the average number of packets in the whole network at each time step. Therefore we can also range the nodes by $h_{p_i}(0)/w_i$. As the complexity of solving a system of linear equations with $n$ variables is $O(n^3)$, the distribution strategy can be get at most $O(n^3)$ steps by numerical computation. Further more, with a scale-free (power-law) node-degree distribution \cite{barabasi1999emergence}, most of the real networks, especially communication networks, are sparse. And there are more efficient methods for these sparse networks \cite{tinney1967direct}.

Next we solve entropy rate as a function of detention probability for two simple and basic systems analytically explicitly: (\romannumeral1) a star-like system and (\romannumeral2) a chain-like system \cite{gao2011robustness}. For simplicity, we suppose the detention probability of node $N_i$ is $q_i, 0\leq q_i<1$ and $q=(q_1,\dots,q_n)$. (\romannumeral1) For a starlike system composed from $n$ nodes as figure 1(a) shows, we have the transition matrix $M_s=(q_{ij})$, where $q_{ii}=q_i, q_{1j}=(1-q_1)/(n-1), q_{j1}=1-q_j, \forall i,j$ and otherwise $q_{ij}=0$. Solve the eigenvector of $M_s$ and we have the stationary distribution: $w=(w_i)$ where $w_1=\theta/(1-q_1)$, $w_j=\theta/(n-1)(1-q_n)\forall j\geq2$ and the normalization factor $\theta=1/(1-q_1)+\sum_{j=2}^{n}1/(n-1)(1-q_j)$. Calculate entropy rate from equations \ref{eq:station} and \ref{eq:entropy}:
\begin{equation}\label{eq:hstar}
  h(q)=\frac{\frac{q_1\ln q_1}{1-q_1}+\ln\frac{1-q_1}{n-1}+\sum_{i=2}^{n}\frac{(1-q_i)\ln(1-q_i)+q_i\ln q_i}{(n-1)(1-q_i)}}{\frac{1}{1-q_1}+\sum_{k=2}^{n}\frac{1}{(n-1)(1-q_k)}}.
\end{equation}
(\romannumeral2) For a chain-like system composed from $n$ nodes as figure 1(b) shows, we suppose the transition matrix $M_c=(q_{ij})$, where $q_{11}=q_1,q_{12}=1-q_2,q_{nn-1}=1-q_n,q_{nn}=q_n$ and $q_{ii-1}=q_{ii+1}=(1-q_i)/2, q_{ii}=q_i, \forall 2\leq i\leq n-1$, otherwise $q_{ij}=0$. Similar to the above, we get the stationary distribution $w=(w_i)$, where $w_j=\mu/(1-q_j)$ for $j=1,n$, $w_j=2\mu/(1-q_j)$ for $2\leq j\leq n$, and the normalization factor $\mu=\sum_{i=2}^{n-1}2/(1-q_i)+\sum_{j=1,n}1/(1-q_j)$. Then the entropy rate is
\begin{equation}\label{hchain}
  h(q)=\frac{\sum_{j=1,n}\frac{q_j\ln q_j+(1-q_j)\ln(1-q_j)}{1-q_j}+\sum_{i=2}^{n-1}\frac{2q_i\ln q_i+(1-q_i)\ln\frac{1-q_i}{2}}{1-q_i}}{\sum_{i=2}^{n-1}\frac{2}{1-q_i}+\sum_{j=1,n}\frac{1}{1-q_j}}.
\end{equation}
For these two systems, we can analyze their entropy rate functions in the neighborhood of dentation probability $(q_{11},\dots,q_{nn})$ and apply our method.

\begin{figure}
  \centering
  \subfigure[]{
  \label{vctree}
  \includegraphics[scale=0.28]{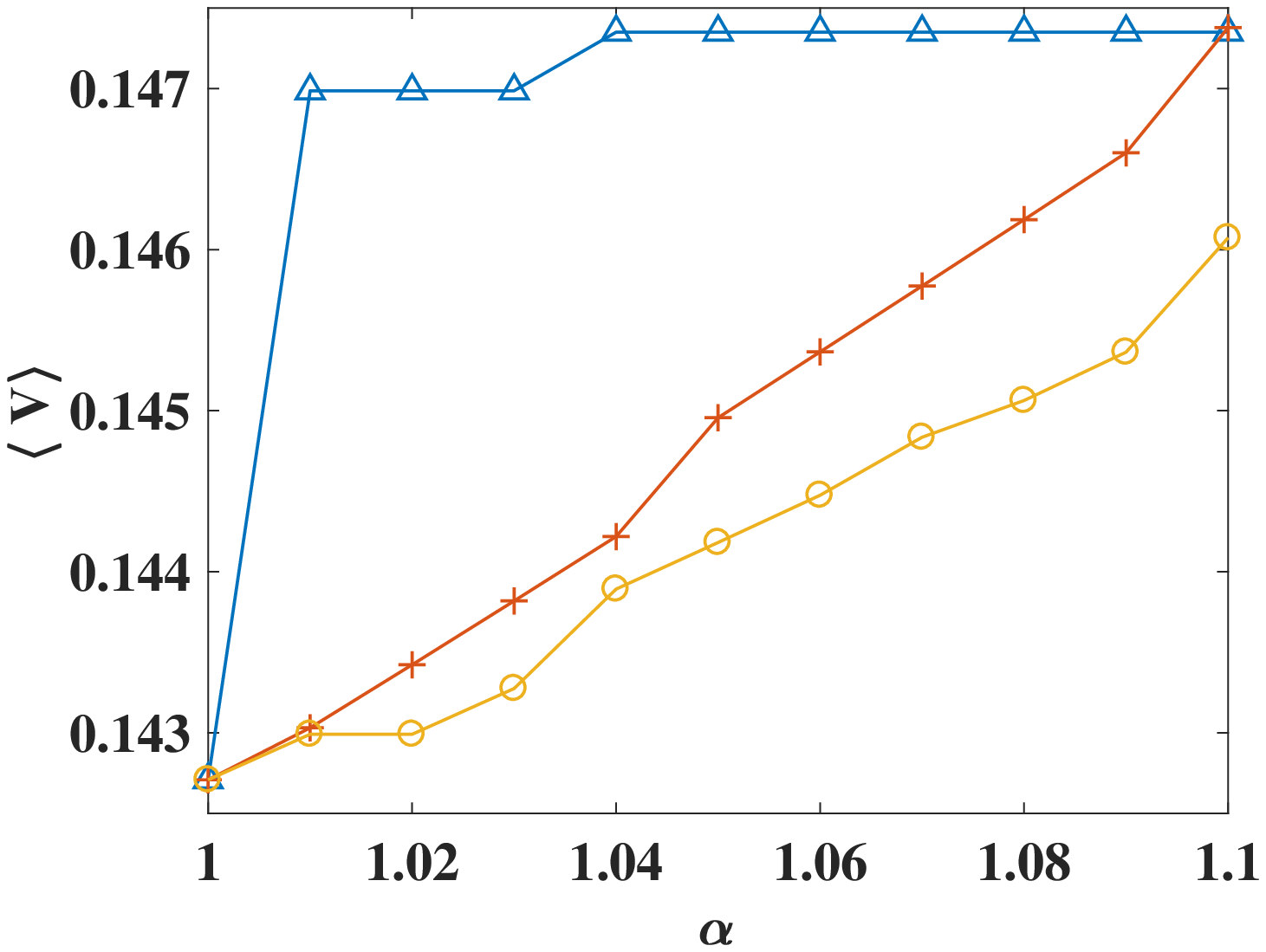}}
  \subfigure[]{
  \label{vregular}
  \includegraphics[scale=0.28]{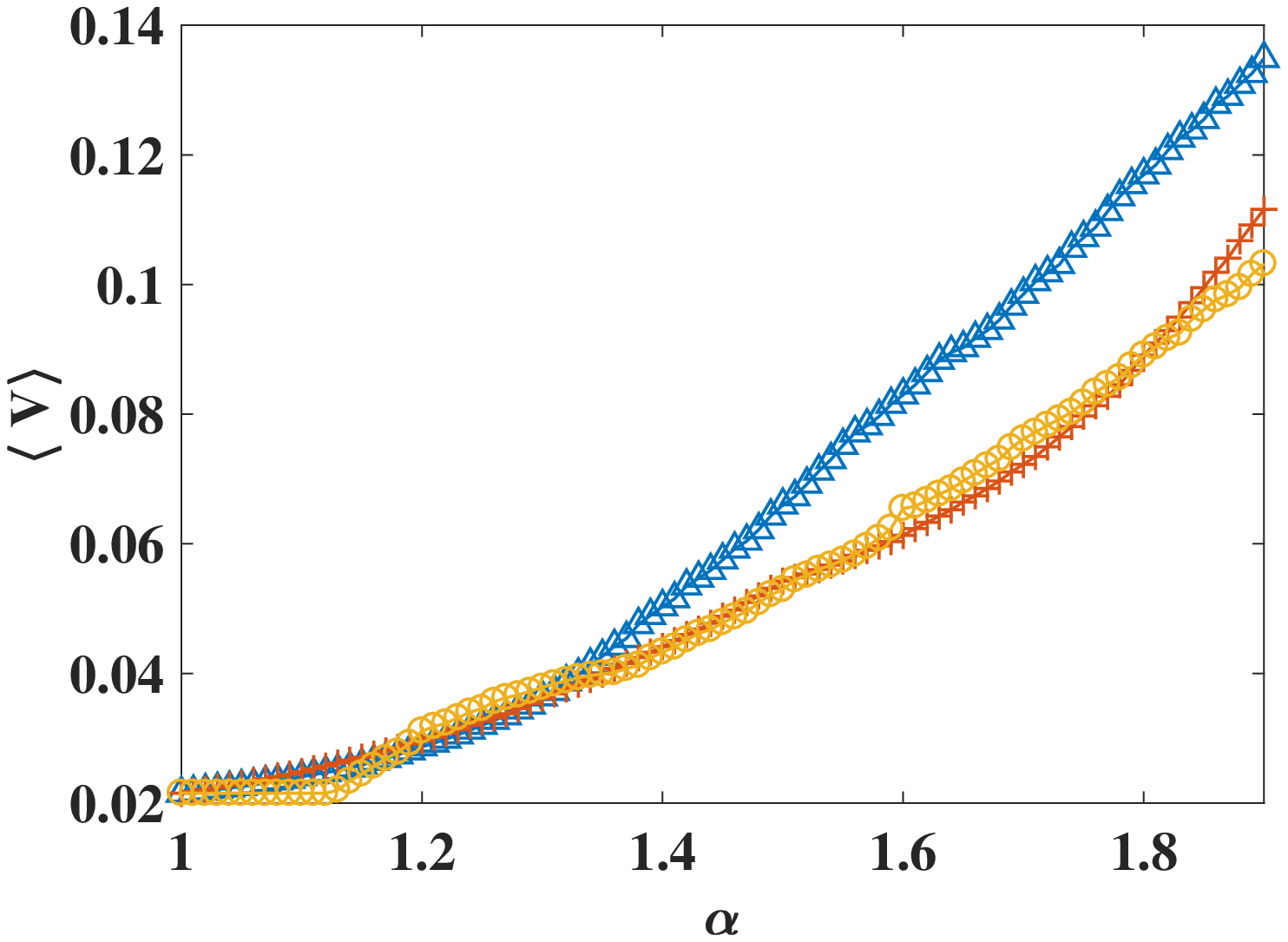}}
  \subfigure[]{
  \label{vrandom}
  \includegraphics[scale=0.28]{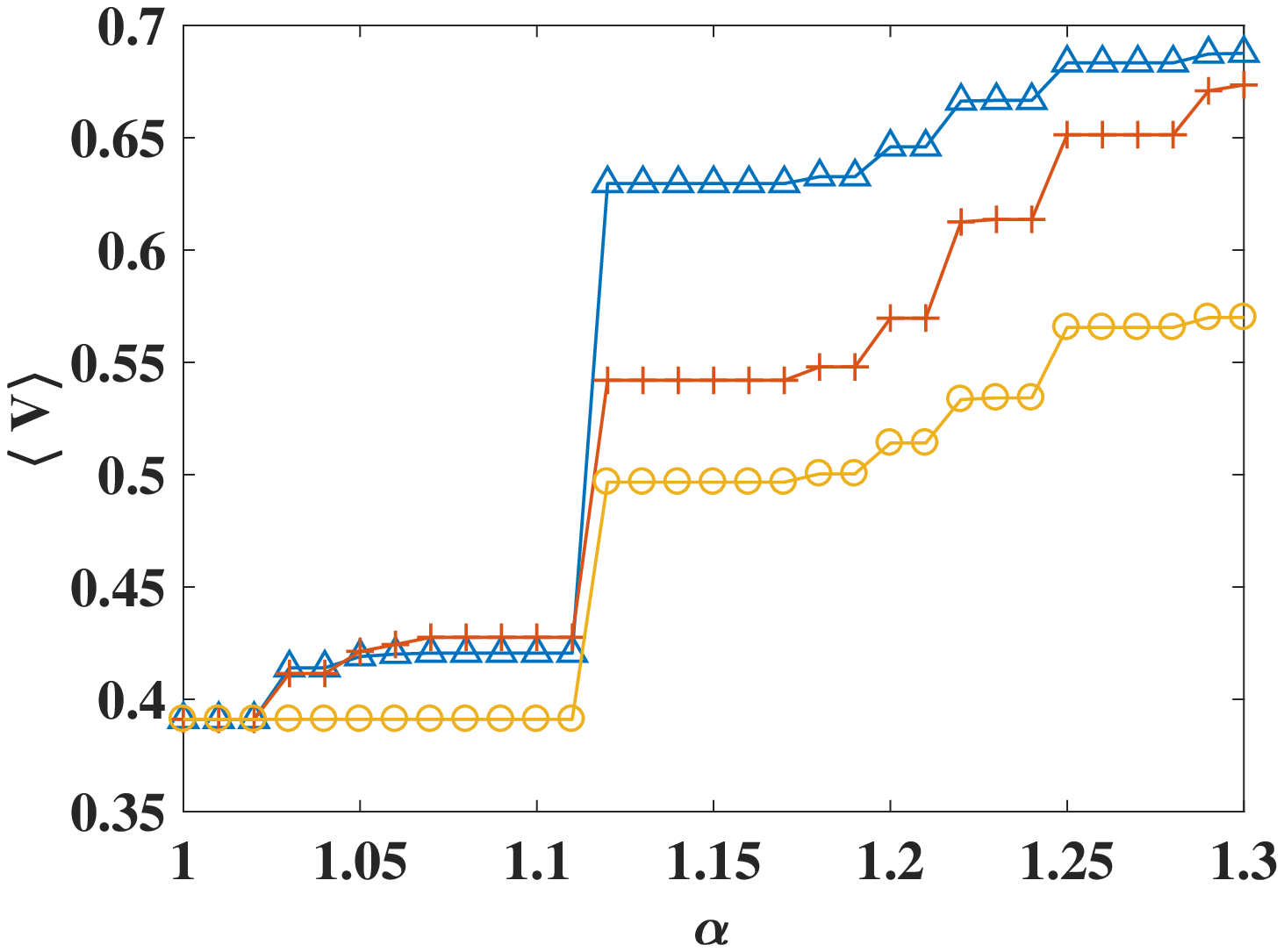}}
  \subfigure[]{
  \label{vscale}
  \includegraphics[scale=0.28]{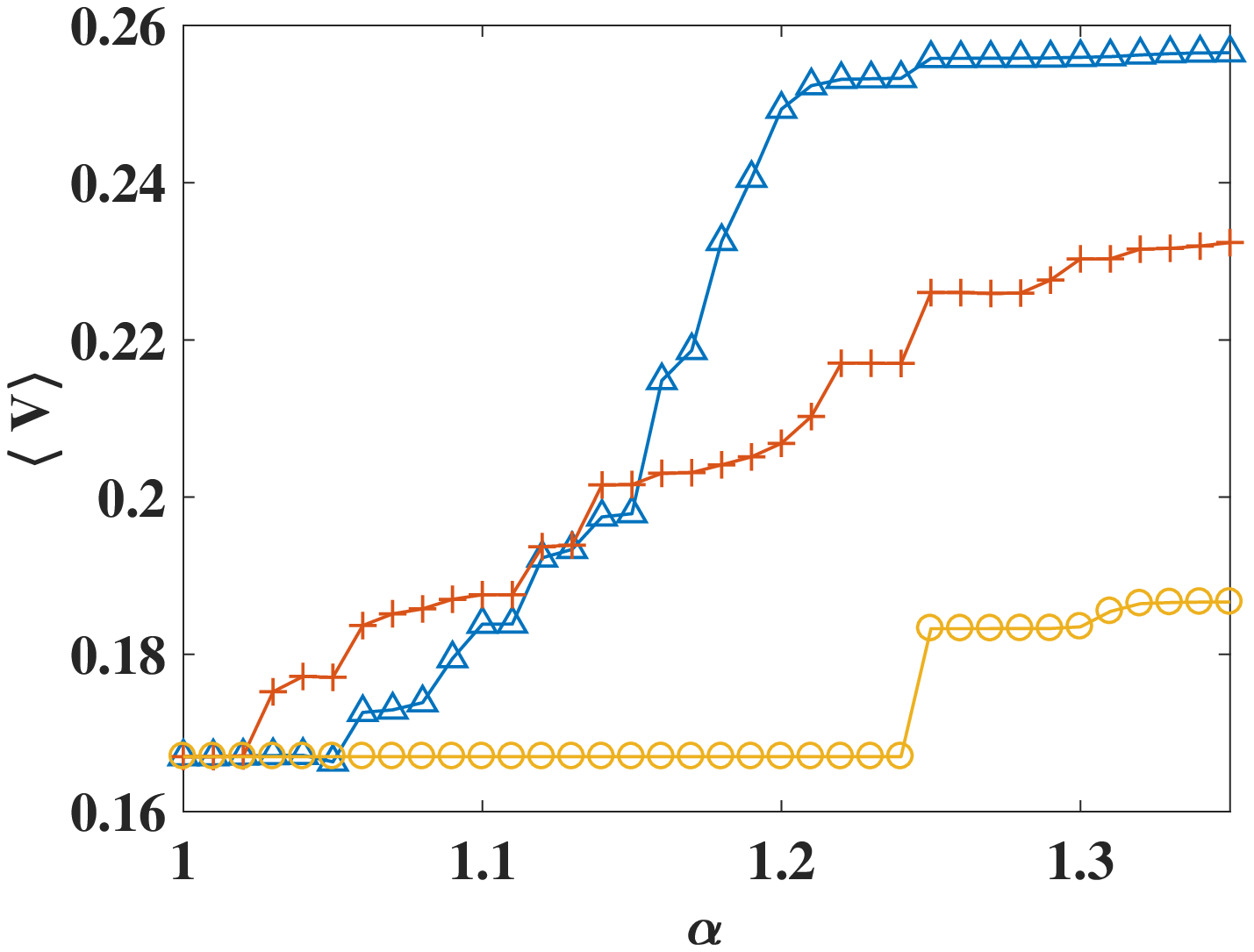}}
  \caption{(Colour On-line) $\langle V\rangle-\alpha$ graph, the average speed $\langle V\rangle$ is shown as a function of $\alpha$ for (a) Cayley tree, $z=3$, $l=6$, $\langle k\rangle=2$, thus $N=1093$, $\beta=30$ (b) regular network, $N=1000$, $\langle k\rangle=4$, $\beta=0.4$, (c) random network, $N=1000$, $\langle k\rangle=4$, $\beta=0.3$ (d) scale-free network, $N=1000$, $\beta=0.2$, with different capacity allocation methods, where the argument $\alpha>1$ is the ratio between initial capacity sum and improved capacity sum of the whole network, and triangle, plus and circle curves correspond to the simulations of our method, degree-biased method and random method. The initial capacity distribution follows $C_i=1+int[\beta k_i]$ at the beginning. We simulate the process of congestion by keeping $\lambda=0.8$ for $500$ time steps and choosing the generating packages' destinations randomly.}\label{va}
\end{figure}

We test our method for the congestion model based on reference \cite{zhao2005onset}, where the node capacity $C_i=1+int[\beta k_i]$ at the beginning, $0\leq\beta\leq1$ is a control parameter and $k_i$ is the degree of node $N_i$. The total number of nodes in the network is $n$. As $\lambda$, the possibility for a node to generate a package at each time step, is increased from zero, two phases can be observed: free flow for small $\lambda$ and a congested phase for large $\lambda$. And there is a phase transition at $\lambda_c$. To simulate the practical congestion process, such as the traffic congestion during peak hours and the congestion of communication during major holiday, we create $\lambda n$ packets at each time step with some probability distribution for a time period, and stop creating after congestion occurring for a while, then the congestion cease after a time period. As what we have mentioned before, we count the frequency in the process of congestion and apply our method to allocate limited capacity budget to the nodes in network. It is worth mentioned that, different from previous work \cite{zhao2005onset}, our method and calculation, based on observation, do not require a known probability distribution of packet generation and the packets' concrete routing strategy. For comparison, we allocate the same capacity budget randomly and by a degree-biased strategy, where the degree-biased strategy, similar to reference \cite{zhao2005onset}, is to increase the control parameter $\beta$ to allocate capacity budget to nodes. Then we create packets in the same way again and compare the traffic efficiency between these methods. Here we use quantity $\langle V\rangle$ to characterize the traffic efficiency. Here $V$ is the velocity of a node being navigating in network.
\begin{equation}\label{velocity}
  V=\frac{L}{T},
\end{equation}
where $L$ is the length (i.e. the number of nodes a packets passed from start position to destination) and $T$ is sum of time steps it takes. The $\langle\dots\rangle$ indicates the average over all packets. We present our simulation for (\romannumeral1) Cayley tree, (\romannumeral2) regular network, (\romannumeral3) random network and (\romannumeral4) scale-free network in Fig. \ref{va}. As it shows, we can see that, for different node capacity budgets, corresponding $\langle V\rangle$ being lower than others in general, our method generally performs well in improving the traffic efficiency with limited capacity budget.

\begin{figure}
  \centering
  \subfigure[]{
  \label{drandom}
  \includegraphics[scale=0.4]{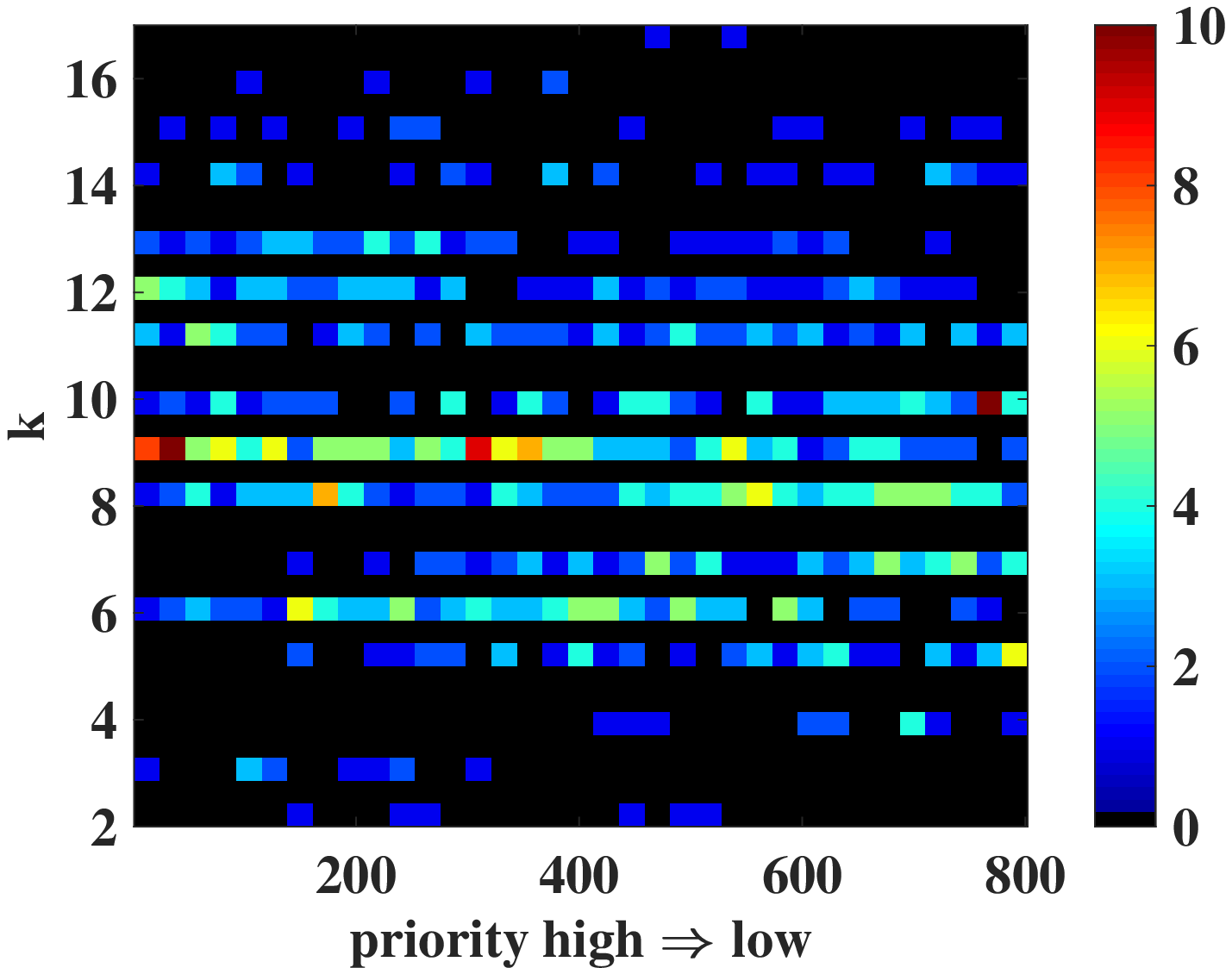}}
  \subfigure[]{
  \label{dscale}
  \includegraphics[scale=0.4]{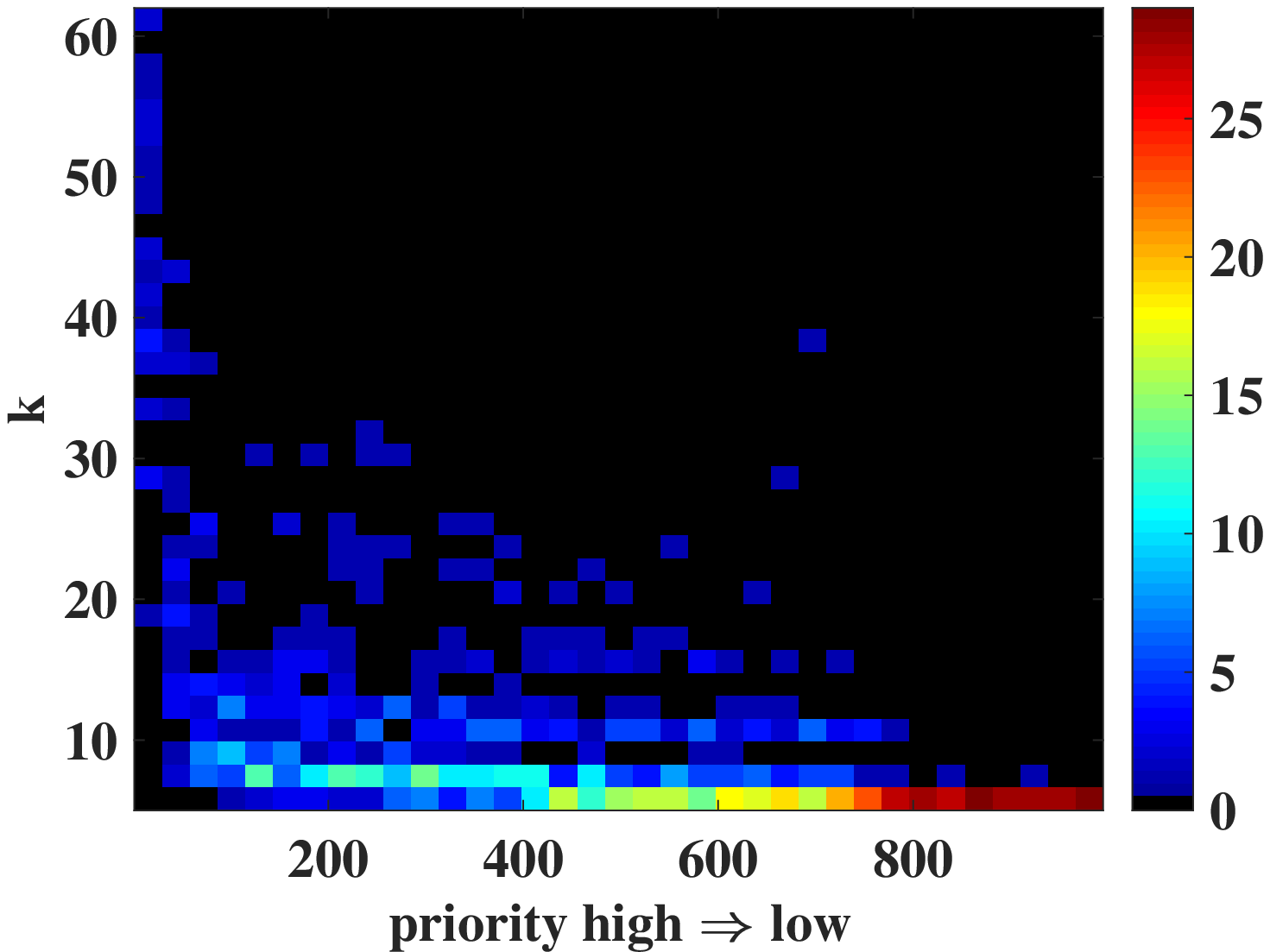}}
  \caption{(Colour On-line) Our method identifies the important nodes for limited capacity budget more reliably than the degree-biased method. The nodes are permutated with their priority $p=h_{p_i}(0)/Q_i$ computed from the equation \ref{hpartial} in the order from highest to lowest on the horizontal axis and the ordinate axis corresponds to their degree $\langle k\rangle$. The distribution of $(p,\langle k\rangle)$ is shown via heat graph for (a) random network and (b) scale-free network. The conditions where we get the results from are the same as the conditions in Fig. \ref{va} for random network and scale-free network.}{\label{degree}}
\end{figure}

Former approaches of controlling traffic congestion, especially for heterogenous networks such as scale-free networks \cite{liu2007method,liu2006efficient}, concentrated on optimizing the nodes with high degree. However, as Fig. \ref{degree} shows, we find that the importance of nodes in the traffic process is not totaly dependent on the their degree, that is, the nodes with high degree may not more important and need to be allocated more capacity than those with low degree. The nodes with higher priority are fairly independent of their degree. For a fixed degree range, there is a wide spread of priority. Therefore our method might be a modification of the former degree-biased study.

\begin{figure}
  \centering
  \subfigure[]{
  \label{dcrandom}
  \includegraphics[scale=0.28]{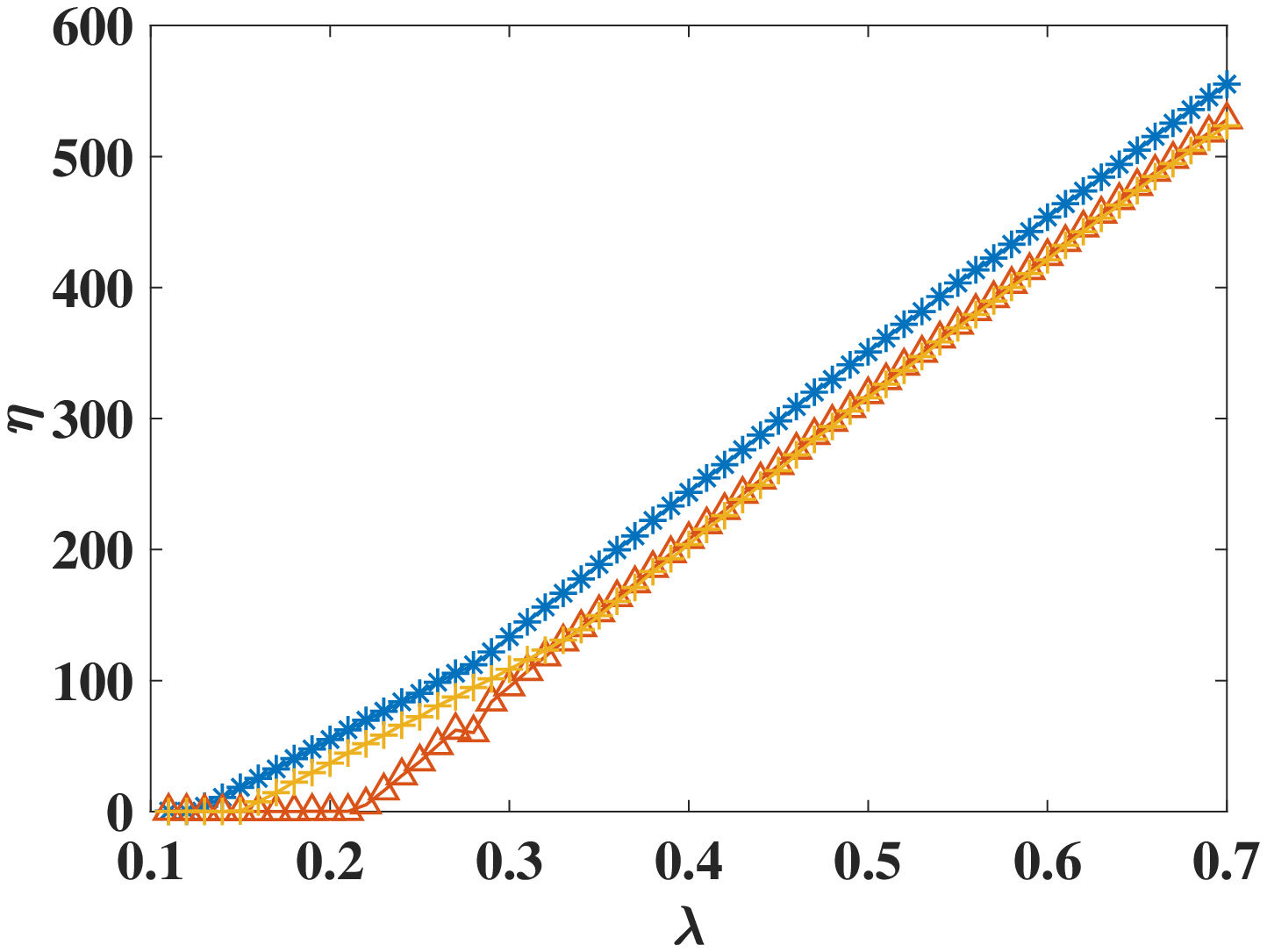}}
  \subfigure[]{
  \label{dscale}
  \includegraphics[scale=0.28]{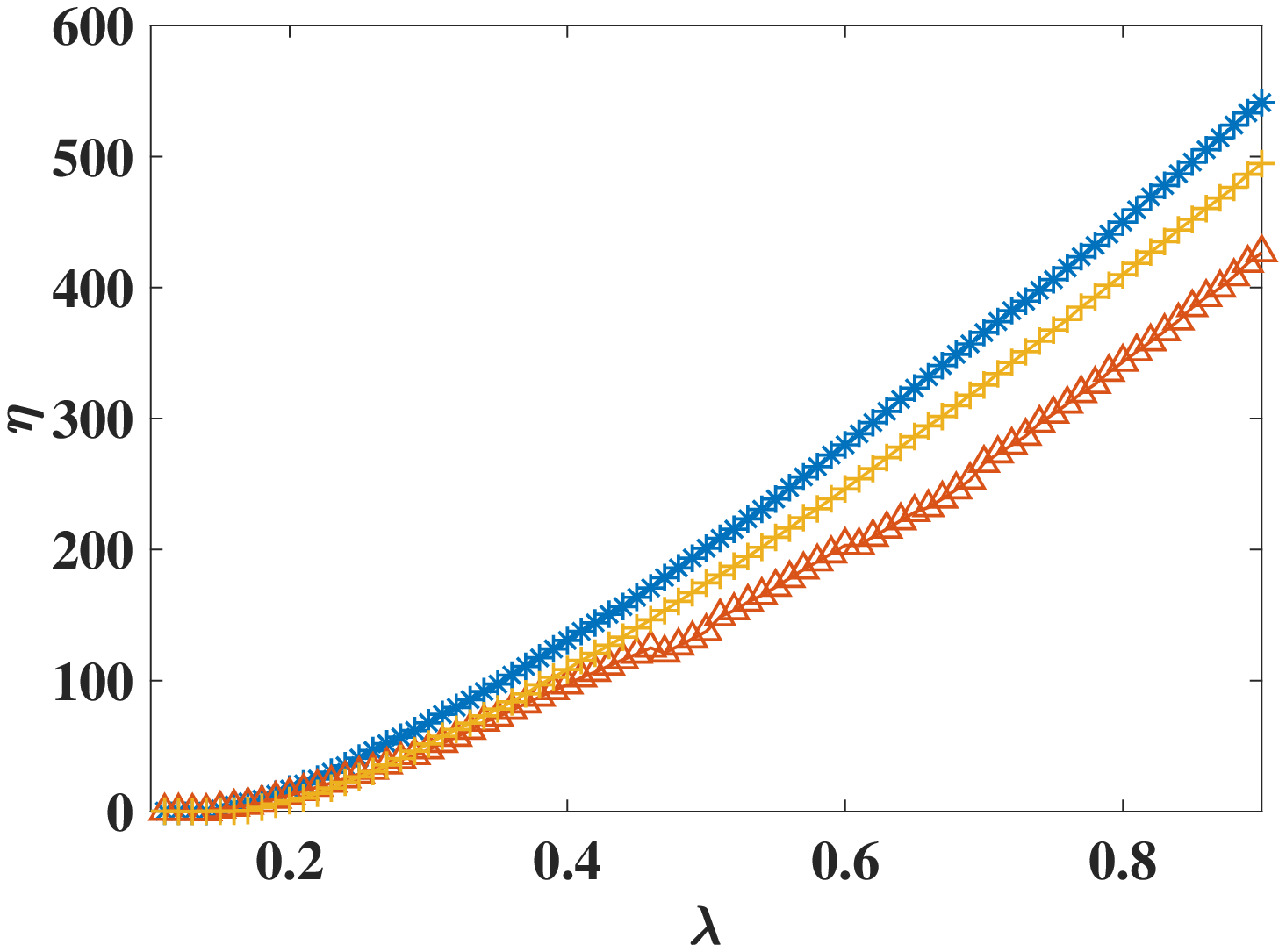}}
  \caption{(Colour On-line) $\eta-p$ graph, the parameter $\eta$ is shown as a function of $\lambda$ for (a) Cayley tree, $z=3$, $l=6$, $\langle k\rangle=2$, $N=1093$, $\beta=30$, $\alpha=0.2$, and (b) scale-free network, $N=1000$, $\beta=0.2$, $\alpha=1.2$, with different optimization methods, where triangle, plus, star curves correspond to the simulations of our method, degree-biased method and original situation.}{\label{figeta}}
\end{figure}

Further more, we reveal that our method can mitigating the congestion globally with local information. In more detail, the transition frequency of each node what we get is the phase of network in a special phase with a $\lambda$. However, as the stationary distribution and entropy rate are global for the dynamical process on complex network, our method may also be effectual on a much larger scale. By simulation, we find that we can optimize a wide range of phases with only the information got in one phase. We allocate capacity budget by our method for a proper phase with a large $\lambda$. And then test the network with different $\lambda$. Similar to the above, we repeat the same process for random strategy and degree-biased strategy as a comparison. It has been showed that \cite{arenas2001communication}, after congestion occurring, the number of packets in the network $Q$ will grow linearly with time. And there is a parameter:
\begin{equation}\label{eta}
  \eta=\lim_{t\to\infty}\frac{\langle\Delta Q\rangle}{\lambda\Delta t}
\end{equation}
to qualify the transition, where $\Delta Q=Q(t+1)-Q(t)$ and $\langle\dots\rangle$ indicates the average over time windows of $\Delta t$. We use this parameter to evaluate the global effect of our optimization for two heterogenous networks: Cayley tree and scale-free network. As Fig. \ref{figeta} shows, the corresponding $\eta$ of method is lower and the threshold of phase transition is higher than others. Our method has a global optimal effect on mitigating congestion.

Introducing maximum entropy rate method into traffic network may also help us assess a traffic network globally and evaluate the gap between it and the optimal state by comparing it to the maximal-entropy random walk \cite{burda2009localization,sinatra2011maximal}, which can be relatively localized easily. Suppose that the practical transition matrix is $M=(p_{ij})$, and the transition matrix corresponding to maximal-entropy random walk is $\Pi={\pi_{ij}}$. Regarding the transition matrix as random variables, we can use Kullback-Leibler divergence to identify the difference between them:
\begin{equation}\label{Kull}
  D(M\|\Pi)=\sum_{\pi_{ij}\neq0}p_{ij}\ln\frac{p_{ij}}{\pi_{ij}}.
\end{equation}
$D(M\|\Pi)\geq 0$ and $D(M\|\Pi)=0$ if and only if $M=\Pi$. We can conclude that a traffic process with lower Kullback–Leibler divergence is more efficient generally.

Recently, it has been showed that many real networks are temporal networks which has many advantages\cite{li2016fundamental}. As our method, based on the dynamical state of network, is less dependent on the fixed network topology. It maybe suitable and helpful for the optimization of the communication on temporal network since the priority can be computed for every time step. Further more, as the capacity usually changes over time in temporal networks too, our method may enlighten researchers to study how to allocate traffic capacity dynamically on temporal networks to improving its traffic efficiency.

Summarizing, in this paper, we analyze the quantitative relationship between traffic capacity in congestion and entropy rate in complex network and solve it explicitly for star-like system and chain-like system. Based on the principle that a higher entropy rate leads to the easiness of propagating in complex network, we develop a method to mitigating congestion by allocating the limited traffic capacity to nodes rationally. We analyze the complexity of our method, get that it can be completed in at most $O(n^3)$ steps, therefore is feasible. What is more, as it only requires the local and observable information and does not change the topology of network, this method is shown to be low-cost and generally practical. We test our method and compare it to other methods for (\romannumeral1) Cayley tree, (\romannumeral2) regular network, (\romannumeral3) random network and (\romannumeral4) scale-free network amd get that our method is more functional in general. The congestion is considerably mitigated after applying our strategy. Free flows on networks are critical in our world and how to mitigate congestion with limited budget is a fundamental and significant problem. Our proposed a heuristic method for this problem and may be useful for modern society.

\bibliographystyle{unsrt}%
\bibliography{mycite_maximum}

\begin{thebibliography}{10}

\bibitem{shannon2001mathematical}
Claude~Elwood Shannon.
\newblock A mathematical theory of communication.
\newblock {\em ACM SIGMOBILE Mobile Computing and Communications Review},
  5(1):3--55, 2001.

\bibitem{reichl1980modern}
Linda~E Reichl and Ilya Prigogine.
\newblock {\em A modern course in statistical physics}, volume~71.
\newblock University of Texas press Austin, 1980.

\bibitem{rosvall2005networks}
Martin Rosvall, Ala Trusina, Petter Minnhagen, and Kim Sneppen.
\newblock Networks and cities: An information perspective.
\newblock {\em Physical Review Letters}, 94(2):028701, 2005.

\bibitem{bianconi2009entropy}
Ginestra Bianconi.
\newblock Entropy of network ensembles.
\newblock {\em Physical Review E}, 79(3):036114, 2009.

\bibitem{gomez2008entropy}
Jes{\'u}s G{\'o}mez-Garde{\~n}es and Vito Latora.
\newblock Entropy rate of diffusion processes on complex networks.
\newblock {\em Physical Review E}, 78(6):065102, 2008.

\bibitem{park2004statistical}
Juyong Park and Mark~EJ Newman.
\newblock Statistical mechanics of networks.
\newblock {\em Physical Review E}, 70(6):066117, 2004.

\bibitem{burda2009localization}
Zdzis{\l}aw Burda, J~Duda, JM~Luck, and B~Waclaw.
\newblock Localization of the maximal entropy random walk.
\newblock {\em Physical review letters}, 102(16):160602, 2009.

\bibitem{arenas2001communication}
Alex Arenas, Albert D{\'\i}az-Guilera, and Roger Guimera.
\newblock Communication in networks with hierarchical branching.
\newblock {\em Physical Review Letters}, 86(14):3196, 2001.

\bibitem{li2015percolation}
Daqing Li, Bowen Fu, Yunpeng Wang, Guangquan Lu, Yehiel Berezin, H~Eugene
  Stanley, and Shlomo Havlin.
\newblock Percolation transition in dynamical traffic network with evolving
  critical bottlenecks.
\newblock {\em Proceedings of the National Academy of Sciences},
  112(3):669--672, 2015.

\bibitem{sole2016congestion}
Albert Sol{\'e}-Ribalta, Sergio G{\'o}mez, and Alex Arenas.
\newblock Congestion induced by the structure of multiplex networks.
\newblock {\em Physical review letters}, 116(10):108701, 2016.

\bibitem{zhao2005onset}
Liang Zhao, Ying-Cheng Lai, Kwangho Park, and Nong Ye.
\newblock Onset of traffic congestion in complex networks.
\newblock {\em Physical Review E}, 71(2):026125, 2005.

\bibitem{eisler2005random}
Zolt{\'a}n Eisler and J{\'a}nos Kert{\'e}sz.
\newblock Random walks on complex networks with inhomogeneous impact.
\newblock {\em Physical Review E}, 71(5):057104, 2005.

\bibitem{goh2001universal}
K-I Goh, B~Kahng, and D~Kim.
\newblock Universal behavior of load distribution in scale-free networks.
\newblock {\em Physical Review Letters}, 87(27):278701, 2001.

\bibitem{liu2007method}
Zhe Liu, Mao-Bin Hu, Rui Jiang, Wen-Xu Wang, and Qing-Song Wu.
\newblock Method to enhance traffic capacity for scale-free networks.
\newblock {\em Physical Review E}, 76(3):037101, 2007.

\bibitem{guimera2002optimal}
Roger Guimer{\`a}, Albert D{\'\i}az-Guilera, Fernando Vega-Redondo, Antonio
  Cabrales, and Alex Arenas.
\newblock Optimal network topologies for local search with congestion.
\newblock {\em Physical Review Letters}, 89(24):248701, 2002.

\bibitem{kleinberg2000navigation}
Jon~M Kleinberg.
\newblock Navigation in a small world.
\newblock {\em Nature}, 406(6798):845--845, 2000.

\bibitem{berger1996maximum}
Adam~L Berger, Vincent J~Della Pietra, and Stephen A~Della Pietra.
\newblock A maximum entropy approach to natural language processing.
\newblock {\em Computational linguistics}, 22(1):39--71, 1996.

\bibitem{bohn2007structure}
Steffen Bohn and Marcelo~O Magnasco.
\newblock Structure, scaling, and phase transition in the optimal transport
  network.
\newblock {\em Physical review letters}, 98(8):088702, 2007.

\bibitem{boyd2004convex}
Stephen Boyd and Lieven Vandenberghe.
\newblock {\em Convex optimization}.
\newblock Cambridge university press, 2004.

\bibitem{barabasi1999emergence}
Albert-L{\'a}szl{\'o} Barab{\'a}si and R{\'e}ka Albert.
\newblock Emergence of scaling in random networks.
\newblock {\em Science}, 286(5439):509--512, 1999.

\bibitem{tinney1967direct}
William~F Tinney and John~W Walker.
\newblock Direct solutions of sparse network equations by optimally ordered
  triangular factorization.
\newblock {\em Proceedings of the IEEE}, 55(11):1801--1809, 1967.

\bibitem{gao2011robustness}
Jianxi Gao, Sergey~V Buldyrev, Shlomo Havlin, and H~Eugene Stanley.
\newblock Robustness of a network of networks.
\newblock {\em Physical Review Letters}, 107(19):195701, 2011.

\bibitem{liu2006efficient}
Zonghua Liu, Weichuan Ma, Huan Zhang, Yin Sun, and Pak~Ming Hui.
\newblock An efficient approach of controlling traffic congestion in scale-free
  networks.
\newblock {\em Physica A: Statistical Mechanics and its Applications},
  370(2):843--853, 2006.

\bibitem{sinatra2011maximal}
Roberta Sinatra, Jes{\'u}s G{\'o}mez-Gardenes, Renaud Lambiotte, Vincenzo
  Nicosia, and Vito Latora.
\newblock Maximal-entropy random walks in complex networks with limited
  information.
\newblock {\em Physical Review E}, 83(3):030103, 2011.

\bibitem{li2016fundamental}
Aming Li, Sean~P Cornelius, Yang-Yu Liu, Long Wang, and Albert-L{\'a}szl{\'o}
  Barab{\'a}si.
\newblock The fundamental advantages of temporal networks.
\newblock {\em arXiv preprint arXiv:1607.06168}, 2016.

\end{thebibliography}
\end{document}